\documentclass[prb,twocolumn,showpacs,showkeys,preprintnumbers,amsmath,amssymb,superscriptaddress]{revtex4}

\usepackage{graphicx}
\usepackage{dcolumn}
\usepackage{bm}

\newcommand{\bra}[1]{\ensuremath{\left\langle #1\right|}}
\newcommand{\ket}[1]{\ensuremath{\left|#1\right\rangle}}
\newcommand{\braket}[2]{\ensuremath{\left\langle #1\vphantom{#2}\right.\left|\vphantom{#1}#2\right\rangle}}
\newcommand{\mean}[1]{\ensuremath{\left\langle #1\right\rangle}}

\begin{document}

\title{Rabi spectroscopy of a qubit-fluctuator system}

\author{J\"urgen Lisenfeld}
\affiliation{Physikalisches Institut, Karlsruhe Institute of Technology,
D-76128 Karlsruhe, Germany}

\author{Clemens M\"uller}
\affiliation{Institut f\"ur Theorie der Kondensierten Materie,
Karlsruhe Institute of Technology, D-76128 Karlsruhe, Germany}
\affiliation{DFG-Center for Functional Nanostructures (CFN), D-76128 Karlsruhe, Germany}

\author{Jared H. Cole}
\affiliation{Institut f\"ur Theoretische Festk\"orperphysik,
Karlsruhe Institute of Technology, D-76128 Karlsruhe, Germany}
\affiliation{DFG-Center for Functional Nanostructures (CFN), D-76128 Karlsruhe, Germany}

\author{Pavel Bushev}
\affiliation{Physikalisches Institut, Karlsruhe Institute of Technology,
D-76128 Karlsruhe, Germany}
\affiliation{DFG-Center for Functional Nanostructures (CFN), D-76128 Karlsruhe, Germany}

\author{Alexander Lukashenko}
\affiliation{Physikalisches Institut, Karlsruhe Institute of Technology,
D-76128 Karlsruhe, Germany}

\author{Alexander Shnirman}
\affiliation{Institut f\"ur Theorie der Kondensierten Materie,
Karlsruhe Institute of Technology, D-76128 Karlsruhe, Germany}
\affiliation{DFG-Center for Functional Nanostructures (CFN), D-76128 Karlsruhe, Germany}

\author{Alexey V. Ustinov}
\affiliation{Physikalisches Institut, Karlsruhe Institute of Technology,
D-76128 Karlsruhe, Germany}
\affiliation{DFG-Center for Functional Nanostructures (CFN), D-76128 Karlsruhe, Germany}

\date{\today}

\begin{abstract}
	Superconducting qubits often show signatures of coherent coupling to
	microscopic two-level fluctuators (TLFs), which manifest themselves
	as avoided level crossings in spectroscopic data. In this work we
	study a phase qubit, in which we induce Rabi oscillations by
	resonant microwave driving. When the qubit is tuned close to
	the resonance with an individual TLF and the Rabi driving is strong enough 
	(Rabi frequency of order of the qubit-TLF coupling), interesting 4-level dynamics are observed.
	The experimental data shows a clear asymmetry between biasing the
	qubit above or below the fluctuator's level-splitting. Theoretical analysis 
	indicates that this asymmetry is due to an effective coupling 
	of the TLF to the external microwave field induced by the higher qubit levels.
\end{abstract}



\maketitle

Spectroscopic analysis of superconducting qubits often shows clear
signatures of avoided level crossings, indicating the presence of microscopic
two-level fluctuators (TLFs) that can be in resonance with the
qubit. Evidence for the existence of TLFs have been found in nearly
all known types of superconducting qubits, including phase-
\cite{Simmonds04,camelback}, flux- \cite{Plourde05,lupascu08},
charge- \cite{kim08}, and transmon qubits \cite{Schreier08}. Since
TLFs are considered to be a source of
decoherence~\cite{Simmonds04,Martinis:2005p1096,Muller:2009}, experiments are
usually conducted by biasing the qubit in a frequency range where
none of these strongly coupled natural two-level systems are
present. Alternatively, one can take advantage of the longer
coherence times of TLFs as compared to the qubits for using them as a quantum memory
\cite{Martinis:2008NatPhys}.
Here we focus on the dynamics of the qubit-fluctuator system on or near resonance.

There are at least two possible mechanisms explaining 
the interaction of the TLFs with the qubit: 
(i) the TLF is an electric dipole which couples to the electric field
in the qubit's Josephson junction~\cite{Martin05,Martinis:2005p1096}.
Nanoscale dipoles could emerge from metastable lattice 
configurations in the amorphous dielectric of the junction's 
tunnel barrier~\cite{Esquinazi};
(ii) the state of the TLF affects the critical current of the qubit's 
Josephson junction~\cite{Simmonds04,deSousa:2009}.
In this case the TLF could be related, e.g., to the formation of Andreev bound states 
at the interface between the superconductor and the 
insulator~\cite{Faoro05,deSousa:2009}.

In this Letter, we explore the complexity of the dynamical  behavior
of a driven phase qubit operated in the vicinity of a
resonance with a two-level fluctuator. Due to the strong coupling
between the qubit and the TLF and equally strong Rabi driving, we observe the dynamics of the resulting
4-level hybrid system consisting of the microscopic defect state and the macroscopic phase qubit. 
Strong microwave driving of the coupled system leads to coherent oscillations, revealing a characteristic 
beating pattern which we analyze quantitatively. Our
experimental data displays a distinct asymmetry of the system response
with respect to the detuning between the qubit and the TLF. 
We argue that this asymmetry is due to Raman-like virtual processes involving 
higher quantum levels of the qubit, giving rise to an effective driving of the TLF.

\begin{figure}[htb]
	\includegraphics[width=\columnwidth]{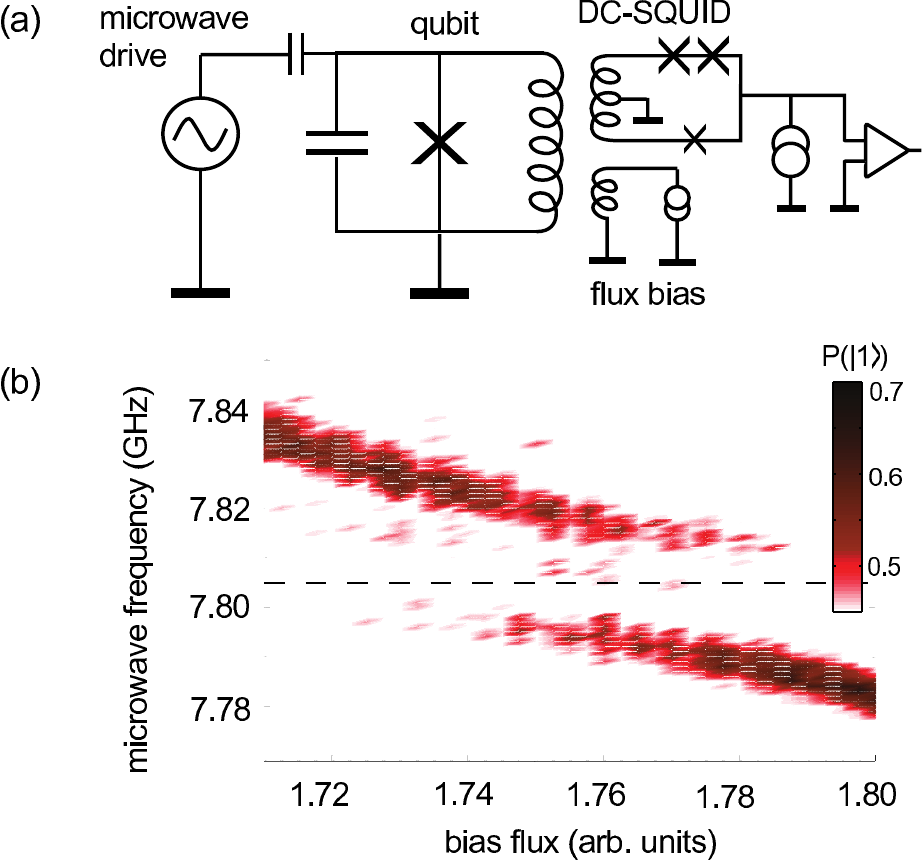}
         \caption{(color online) \textbf{(a)} Schematic of the phase
         		qubit circuit. \textbf{(b)} Probability to measure the excited qubit state (color-coded) vs. bias flux and microwave frequency, 
		revealing the coupling to a two-level defect state having a resonance frequency of 7.805 GHz (indicated by a dashed line). 
	}
	\label{fig:Splitting}
\end{figure}

The sample investigated in this study is a phase
qubit~\cite{Simmonds04}, consisting of a capacitively shunted
Josephson junction embedded in a superconducting loop. 
Its potential energy has the form of a double well for
suitable combinations of the junction's critical current (here, $I_c = 1.1 \mu$A) and loop inductance (here, $L = 720$~pH).
For the qubit states, one uses the two Josephson phase eigenstates of lowest
energy which are localized in the shallower of the two potential
wells, whose depth is controlled by the external magnetic flux through the qubit loop.
The qubit state is controlled by an externally applied microwave pulse, which in our sample is coupled
capacitively to the Josephson junction. A schematic of the complete qubit circuit is depicted in Fig.~\ref{fig:Splitting}(a). 
Details of the experimental setup can be found in Ref.~\onlinecite{LisenfeldPRL07}. 
During all measurements presented in this paper, the sample was cooled to a temperature of 35 mK in a dilution refrigerator.

Spectroscopic data taken in the whole accessible frequency range between 5.8 GHz
and 8.1 GHz showed only 4 avoided level crossings due to TLFs having
a coupling strength larger than 10 MHz, which is about the spectroscopic
resolution given by the linewidth of the qubit transition. 
In this work, we studied the qubit interacting with a
fluctuator whose energy splitting was $\epsilon_f / h = 7.805$~GHz. 
From its spectroscopic signature shown in
Fig.~\ref{fig:Splitting}(b), we extract a coupling strength $v_{\perp} / h \approx 25$~MHz. 
The coherence times of this TLF were measured by directly driving it at its resonance frequency while the qubit was kept detuned\cite{Lisenfeld2010}. A $\pi$ pulse was applied to measure the energy relaxation time $T_{1, f} \approx 850$~ns, while two delayed $\pi/2$ pulses were used to measure the dephasing time $T_{2, f}^{*} \approx 110$~ns in a Ramsey experiment. To read out the resulting TLF state, the qubit was tuned into resonance with the TLF to realize an iSWAP gate, followed by a measurement of the qubit's excited state. 

Experimentally, we observe the probability $P(\ket{e})$ of the
qubit being in its excited state after driving it resonantly with a short
microwave pulse. Varying the duration $\tau$ of the microwave pulse 
allows us to observe the evolution of $P(\ket{e})$ in the time
domain. If the energy splitting of the qubit is tuned far away from
that of the fluctuator, the qubit remains decoupled from the TLF and
$P(\ket{e})$ displays the usual Rabi oscillations in the form of
an exponentially decaying sinusoid having only a single frequency
component. For our qubit sample, which has coherence times of $T_{1,q}\approx 120$~ns 
and $T_{2,q} \approx 90$~ns, these oscillations have the
characteristic decay time of about $115$~ns. 
If, in contrast, the qubit is tuned close to the resonance frequency of a TLF, the
probability to measure the excited qubit state shows a complicated
time dependence, which is very sensitive to the chosen qubit bias.

\begin{figure}[htb]
      	\includegraphics[width=\columnwidth]{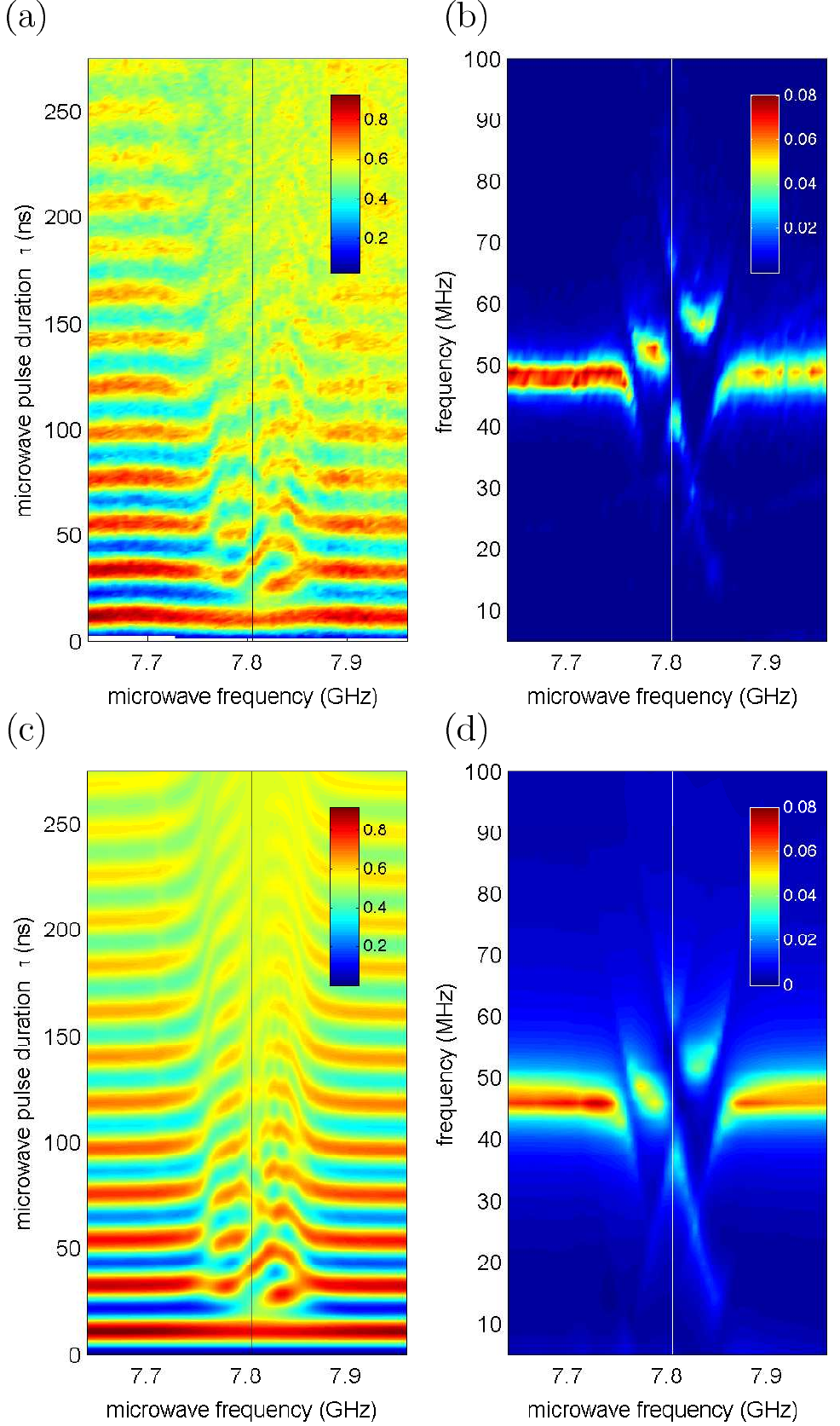}
      	\caption{(color online) \textbf{(a)} Experimentally observed time evolution of the probability
      		to measure the qubit in the excited state, $P(\ket{e})(t)$, vs. driving frequency;
      		\textbf{(b)} Fourier-transform of the experimentally observed $P(\ket{e})(t)$. 
		The resonance frequency of the TLF is indicated by vertical lines. 
     		\textbf{(c)} Time evolution of $P(\ket{e})$ and \textbf{(d)} its Fourier-transform 
		obtained by the numerical solution of Eq.~(\ref{eq:master_eq}) as described in the text, 
		taking into account also the next higher level in the qubit. 
		(As the anharmonicity $\Delta/h \sim 100$~MHz in our circuit is relatively small, 
		this required going beyond the second order perturbation theory 
		and analyze the 6-level dynamics explicitly).
		The qubit's Rabi frequency $\Omega_{q}/h$ is set to 48 MHz. 
		}
      \label{fig:DataRabi}
\end{figure}

Figure~\ref{fig:DataRabi}(a) shows a set of time traces of
$P(\ket{e})$ taken for different microwave drive frequencies. Each
trace was recorded after adjusting the qubit bias to result in an
energy splitting resonant to the chosen microwave frequency. The
Fourier transform of this data, shown in Fig.~\ref{fig:DataRabi}(b), 
allows us to distinguish several frequency components. Frequency
and visibility of each component depend on the detuning between
qubit and TLF. We note a striking asymmetry between the Fourier
components appearing for positive and negative detuning of the qubit
relative to the TLF's resonance frequency, which is indicated in
Figs.~\ref{fig:DataRabi}(a,b) by the vertical lines at 7.805 GHz. 
We argue below that this asymmetry is due to virtual Raman-transitions involving
higher levels in the qubit.

To describe the system theoretically, we write down the Hamiltonian, consisting of two parts: $ \hat{H} = \hat{H}_{S} + \hat{H}_{I} $, 
with $\hat{H}_{S}$ being the system Hamiltonian, representing qubit, TLF and their coupling 
and $\hat{H}_{I}$ describing the interaction between system and microwave driving. 
The Hamiltonian of the qubit circuit reads
\begin{equation}
	H_{S}^{q} = E_{C} \left(n - n_{G} \right)^{2} - E_{J} \cos{\phi} + E_{L} \left( \phi - \phi_{ext} \right)^{2} \,,
\end{equation}
where $E_{C/J/L}$ are charging/Josephson/inductive energies of the circuit, 
$\phi$ is the phase difference across the Josephson junction, and 
$n$ is the dimensionless charge conjugate to $\phi$, i.e., $[\phi,n]=i$.
The circuit can be manipulated by applying an {\it ac} driving to gate charge 
$n_{G}$ or the external flux $\phi_{ext}$.
The TLF is described as a two level system $H_{S}^{f} = 1/2 \epsilon_{f} \tau_{z}$ which couples either to the electric field across the junction $\propto (n-n_{G})$ or, 
alternatively, to the Josephson energy $\propto \cos{\phi}$. The coupling can be 
either transverse, $\propto \tau_{\pm}$, or longitudinal, $\propto \tau_{z}$. 

For maximum generality, we first define a \emph{minimal} model needed to describe the splitting of Fig.~\ref{fig:Splitting}.
To this end, we restrict ourselves to the lowest two states of the phase qubit circuit (the qubit subspace) and disregard the longitudinal 
coupling $\propto \tau_{z}$. Within the rotating wave approximation (RWA) the 
Hamiltonian reads
 \begin{equation}
 	\hat{H}_{S}^{min} =  \frac{1}{2} \epsilon_{q} \sigma_{z} + \frac{1}{2} \epsilon_{f} \tau_{z}
         		+ \frac{1}{2} v_{\perp} \left( \sigma_{-} \tau_{+} + \sigma_{+} \tau_{-} \right) \,,
	\label{eq:HsMin}
\end{equation}
with the Pauli-matrices for the qubit $\sigma$ and for the fluctuator $\tau$.
The minimal interaction Hamiltonian couples only the qubit to the driving field via the coupling constant $\Omega_{q}$:
$\hat{H}_{I}^{min} = \Omega_{q} \cos{(\omega_{d} t)} \: \sigma_{x}$. The RWA is justified 
since $\Omega_q \sim v_\perp \ll \epsilon_q \sim \epsilon_f$.
Rabi oscillations in this minimal system have been considered earlier~\cite{Ashhab:2006p37, Galperin:2007p34}. 

\begin{figure}[htb]
	\includegraphics[width=\columnwidth]{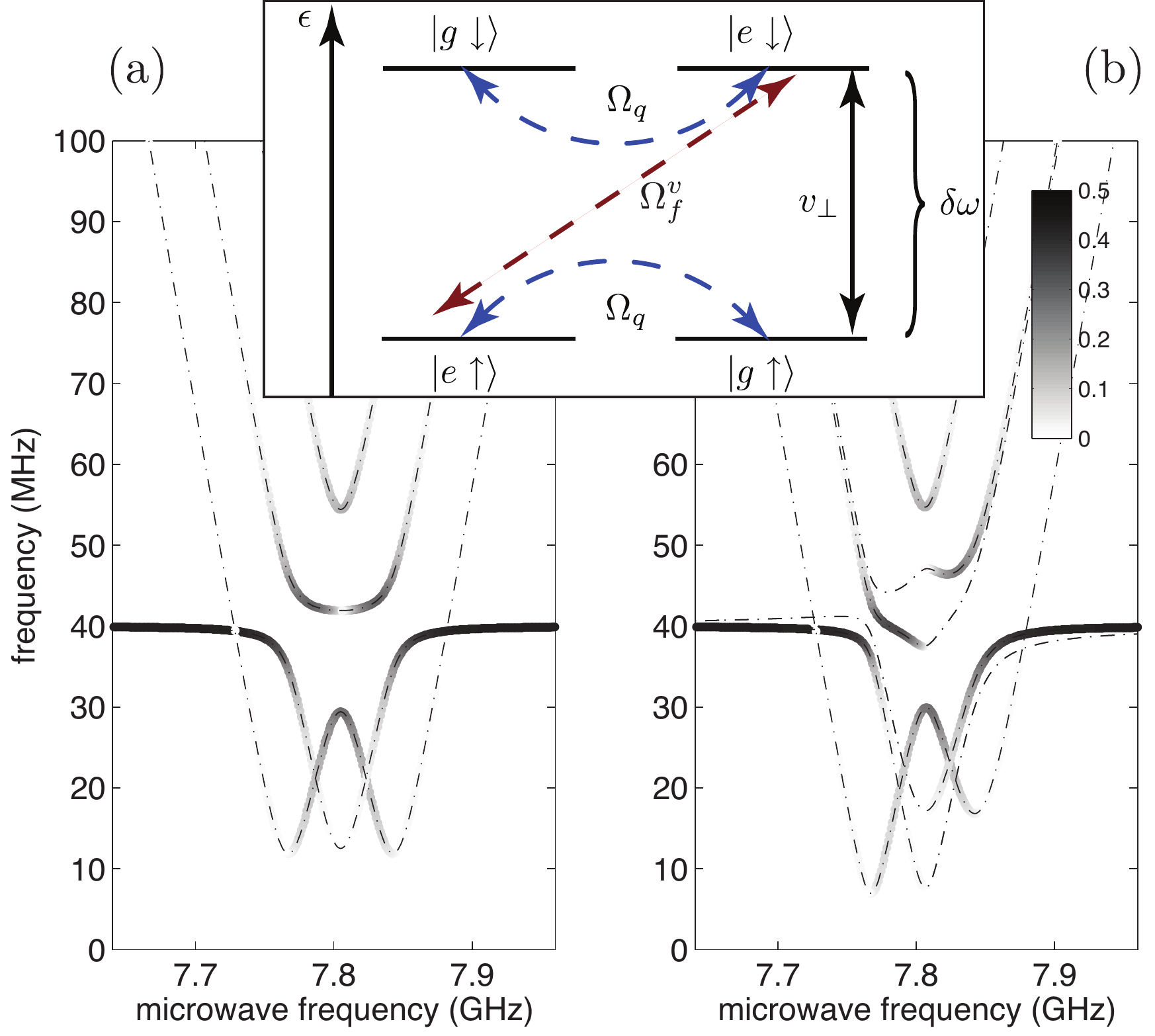}
         \caption{(color online) \textbf{(a)} Analytically obtained transition spectrum of the Hamiltonian (\ref{eq:H4Levels}) in the minimal model
         		for $\Omega_q/h = 40$~MHz and $v_\perp/h = 25$~MHz.
         		Dashed-dotted lines show the transition frequencies while the gray-scale intensity of the thicker 
         		lines indicates the weight of the respective Fourier-components in the probability $P(\ket{e})$.
                	The system shows a symmetric response as a function of the detuning $\delta\omega$.
		Two of the four lines are double degenerate.
                	\textbf{(b)} The same as (a) but including the second order Raman process with 
		$\Omega^{v}_{f} = v_{\perp} \Omega_{q} / \Delta$.
                	The two degenerate transitions in (a) split and the symmetry of the response is broken.
                	\textbf{Inset:} Schematic representation of the structure of the Hamiltonian~(\ref{eq:H4Levels}).
		We denote the ground and excited states of the qubit as $\ket{g}$ and $\ket{e}$ 
		and those of the TLF as $\ket{\downarrow}$ and $\ket{\uparrow}$. 
		Arrows indicate the couplings between qubit and fluctuator $v_{\perp}$ and to the microwave field $\Omega_{q}$ and $\Omega^{v}_{f} $.
	}
	\label{fig:Transitions}
\end{figure}

Going to the rotating frame for both qubit and TLF and taking the frequency of the driving to be resonant with the qubit splitting, $\omega_{d} = \epsilon_{q}$,
we arrive at the effective 4-level Hamiltonian 
\begin{equation}
H^{min} =- \frac{1}{2} \delta\omega \tau_{z}
         		+ \frac{1}{2} v_{\perp} \left( \sigma_{-} \tau_{+} + \sigma_{+} \tau_{-} \right)+
		 \frac{1}{2} \Omega_{q} \sigma_{x}\ ,
	\label{eq:H4Levels}
\end{equation}
where $\delta\omega = \left( \epsilon_{q} - \epsilon_{f} \right)$.
The level structure and the spectrum of possible transitions in the Hamiltonian (\ref{eq:H4Levels}) is illustrated in Fig.~\ref{fig:Transitions}a. 
The transition frequencies in the rotating frame correspond to the frequencies of the Rabi oscillations observed experimentally.

To describe the time evolution of our system we consider the state
$\ket{\Psi (t)} = \sum_{k} c_{k} e^{-i E_{k} t} \ket{k}$,
where $E_{k}$ are the eigenvalues and $\ket{k}$ the eigenstates of the Hamiltonian (\ref{eq:H4Levels}).
The coefficients $c_{k}$ are determined by the initial conditions.
The eigenvectors $\ket{k}$ can be expressed as linear combinations of the measurement basis states $\ket{g\downarrow}$, $\ket{g\uparrow}$, $\ket{e\downarrow}$, $\ket{e\uparrow}$, i.e., the mutual eigenstates of $\sigma_z$ and $\tau_z$, which we denote by 
$\{ \ket{l} \}$ with $l=0,1,2,3$.
For the expectation value of the operator $\sigma_z$ we get
\begin{eqnarray}
	\bra{\Psi} \sigma_z \ket{\Psi} &=& \sum_{k,l,m} a^*_{k,l} a_{m,l} e^{-i (E_{m} - E_{k}) t} \bra{l} \sigma_z\ket{l}\ ,
        	\label{eq:Operator}
\end{eqnarray}
where $a_{k,l} = c_{k} \braket{l}{k}$ and we used the fact that $\sigma_z$ is diagonal in basis $\{ \ket{l} \}$. 
From Eq.~(\ref{eq:Operator}) we can extract the Fourier components of the experimentally measured excited state population 
$P(\ket{e})=(1+\mean{\sigma_z})/2$. 
There are six components with, in general different, transition frequencies $E_{m} - E_{k}$. 
These are shown in Fig. \ref{fig:Transitions}a for the minimal model. Only four lines are seen due to two 
double degeneracies. The intensity of the thick lines overlaying the dashed-dotted transition 
lines corresponds to the amplitude of these Fourier components.
The situation depicted in Fig.~\ref{fig:Transitions} and realized in our experiment corresponds to the 
qubit and the fluctuator initially in their ground states.
It is important to note that the pattern of Fig.~\ref{fig:Transitions} is characteristic for the regime $\Omega_q \sim v_\perp$.

As seen in Fig.~\ref{fig:Transitions}a the observed asymmetry in the response can not be explained by the minimal model. 
We identify three possible mechanisms which could break the symmetry:
(i) Longitudinal coupling between qubit and TLF $H_{S}^{long} \sim v_{\parallel} \sigma_{z} \tau_{z}$. 
We note that the longitudinal coupling is excluded for the electric dipole coupling mechanism in phase and flux qubits, 
since this term would necessitate an average electric field (voltage) across the junction. 
The longitudinal coupling might be present if the TLF couples via a change in the critical current~\cite{Simmonds04,deSousa:2009}. In this case
the state of the TLF directly affects the shape of the Josephson potential, therefore modulating the level-splitting of the qubit.
For realistic parameters, this might lead to a strong longitudinal coupling $v_{\parallel}$.
Such a coupling was, however, ruled out spectroscopically in Ref.~\onlinecite{lupascu08} as well as by our preliminary spectroscopic data~\cite{Bushev2010}. 
(ii) Direct coupling of the TLF to the external field $H_{I}^{d} = \Omega_{f}^{d} \cos{(\omega_{d} t)} \: \tau_{x}$. 
Due to the presumably small size of the TLF this coupling should be negligible. 
(iii) Effective coupling of the TLF to the external driving field due to a second order Raman-like process 
in which the next higher level of the qubit $\ket{e_2}$ is virtually excited followed by a mutual flip 
of the TLF and the qubit (back to state $\ket{e}$). The energy difference between 
the states $\ket{e_2}$ and $\ket{e}$ is given by $\epsilon_q - \Delta$, where 
$\Delta$ characterizes the anharmonicity of the qubit.
This gives an effective coupling $H_{I}^{v} = \Omega_{f}^{v} \cos{(\omega_{d} t)}\: \tau_{x} \ket{e}\bra{e}$, 
i.e., the coupling is present only when the qubit is excited.
For $\delta\omega < \Delta$, we find 
$\Omega_{f}^{v} \approx v_{\perp}\Omega_{q} / \Delta$. 
In Fig \ref{fig:Transitions}b we show the spectrum of transitions with only the term $H_{I}^{v}$ added to the minimal model, Eq.~(\ref{eq:H4Levels}), 
not including longitudinal coupling or direct coupling of the TLF to the driving field.

To fully describe the experiment, we include decoherence in our calculations by solving the time evolution of the
system's density matrix $\rho$ using a standard Lindblad-approach \cite{Gardiner}.
The dynamic equations are given by
\begin{equation}
	\dot{\rho} = i \left[ \rho, \hat{H} \right] + \sum_{j} \Gamma_{j} \left( L_{j} \rho L_{j}^\dagger -\frac{1}{2} \left\{ L_{j}L_{j}^\dagger, \rho \right\} \right)\,,
	\label{eq:master_eq}
\end{equation}
where the sum is over all possible channels of decoherence with the respective rates $\Gamma_{j}$.
The $L_{j}$ are the operators corresponding to each decoherence channel, e.g., pure dephasing of the qubit is described  by the operator $\sigma_{z}$.
The theoretical spectral response of the system obtained by numerically solving the dynamical equations is shown in Fig.~\ref{fig:DataRabi}(c, d). 
Relaxation and pure dephasing rates for qubit and TLF have been taken to be equal to the values mentioned earlier.
The plot of Fig.~\ref{fig:DataRabi}(c,d) takes into account the third level in the qubit.
As the anharmonicity $\Delta/h \sim 100$~MHz is known from other measurements~\cite{Bushev2010}, we have no additional fit parameters and quantitatively reproduce the experimental data. Note that we are able to explain the experimental data by 
assuming $v_{\parallel}=0$, which provides further evidence in favor of 
the dipole coupling mechanism.

In conclusion, we studied the dynamics of a driven system consisting of a phase qubit strongly coupled to a TLF. The Fourier-analysis of the Rabi oscillation data reveals the characteristic pattern of transition frequencies in the coupled system. 
This asymmetric pattern is reproduced quantitatively by the presented theoretical model including virtual transitions to the qubit's higher levels. The apparent absence of the longitudinal coupling between the qubit and the TLF gives a hint about the microscopic nature of the TLFs.

We would like to thank M. Ansmann and J. M. Martinis (UCSB) for providing
us with the sample that we measured in this work. This work was supported
by the CFN of DFG, the EU projects EuroSQIP and MIDAS, and the U.S. ARO under Contract No. W911NF-09-1-0336.

\bibliographystyle{apsrev}
\bibliography{RabiSpectroscopy}

\end{document}